\documentclass[iop]{emulateapj}
\usepackage[usenames]{color}
\usepackage{ulem}
\usepackage{color}
\scshape
\newcommand{\lsun}{$\rm L_{\odot}$ }

\newcommand{\msun}{$\rm M_{\odot}$}

\newcommand{\sigsfr}{$\Sigma_{\rm SFR}$}
\newcommand{\siggas}{$\Sigma_{\rm gas}$}

\newcommand{\msunmyrpc}{$\rm M_{\odot}~{\rm Myr}^{-1}~{\rm pc}^{-2}$}
\newcommand{\msunpc}{$\rm M_{\odot}~{\rm pc}^{-2}$}
\newcommand{\gcm}{$\rm g~{cm}^{-2}$}

\newcommand{\um}{$\mu$m}
\newcommand{\co}{$\rm ^{13}$CO}
\newcommand{\nhdos}{$\rm N_{H_2}$}

\newcommand{\kms}{$\rm km~s^{-1}$}

\newcommand{\irdcs}{{\it IRDCs}}

\date{}
\shorttitle{IRDCs and HMSF activity in GMCs}
\shortauthors{Retes-Romero et al.}

\begin{document}
\title{Infrared Dark Clouds and high-mass star formation activity in Galactic Molecular Clouds}

%% Use \author, \affil, and the \and command to format
%% author and affiliation information.
%% Note that \email has replaced the old \authoremail command
%% from AASTeX v4.0. You can use \email to mark an email address
%% anywhere in the paper, not just in the front matter.
%% As in the title, use \\ to force line breaks.

%\author{R. Retes-Romero\altaffilmark{1} et al.}
\author{R. Retes-Romero\altaffilmark{1,2}, Y. D. Mayya, A. Luna and L. Carrasco}

\affil{Instituto Nacional de Astrof\'isica, \'Optica y Electr\'onica,
Luis Enrique Erro 1, Tonantzintla, Puebla, C.P. 72840, M\'exico.}

\altaffiltext{1}{Email adress: rretes@inaoep.mx, ydm@inaoep.mx, aluna@inaoep.mx, carrasco@inaoep.mx}
\altaffiltext{2}{Current adress: Decanato de Ingenier\'ia, UPAEP, 21 Sur 1103, 
Barrio de Santiago, C.P. 72410, Puebla, M\'exico.}

\begin{abstract}
Ever since their discovery, Infrared dark clouds (IRDCs) are generally 
considered to be the sites just at the onset of high-mass (HM) star formation. 
In recent years, it has been realized that not all IRDCs 
harbour HM Young Stellar Objects (YSOs). Only those IRDCs satisfying a 
certain mass-size criterion, or equivalently above a certain threshold density, 
are found to contain HMYSOs. In all cases, 
IRDCs provide ideal conditions for the formation of stellar clusters. 
In this paper, we study the massive
stellar content of IRDCs to re-address the relation between IRDCs and HM
star formation. For this purpose, we have identified all IRDCs 
associated to a sample of 12 Galactic molecular clouds (MCs).
The selected MCs have been the target of a systematic search for YSOs in
an earlier study. The catalogued positions of YSOs have been used to search 
all YSOs embedded in each identified IRDC. In total, we have found 834 YSOs in 
128 IRDCs. The sample of IRDCs have mean surface densities of 319\,\msunpc,
mean mass of 1062~\msun, and a mass function power-law slope $-1.8$, which are 
similar to the corresponding properties for the full sample of IRDCs and resulting physical 
properties in previous studies.
We find that all those IRDCs containing at least one intermediate to high-mass young star
satisfy the often-used mass-size criterion for forming HM stars.
However, not all IRDCs satisfying the mass-size criterion
contain HM stars. We find that the often used mass-size criterion corresponds to 
35\% probability of an IRDC forming a massive star. Twenty five (20\%) of the IRDCs 
are potential sites of stellar clusters of mass more than 100~\msun.

\end{abstract}

\keywords{ISM: YSOs --- ISM: HII regions --- ISM: infrared dark clouds --- stars: formation}

\section{ Introduction}

Infrared dark clouds (IRDCs) are dark silhouettes observed in the mid-infrared (MIR) 
due to the contrast of the absorption against
the bright emission of the Galactic background in the MIR images first observed using the 
{\it Infrared Space Observatory survey} \citep{perault+96} and the {\it Midcourse Space Experiment} 
\citep{egan+98}. 
These regions are associated to the densest regions of the Galactic molecular 
clouds (MCs) and have physical conditions such as low temperatures 
($\rm T<25~K$) and high densities  ($\rm n\gtrsim10^5~cm^{-3}$) \citep{rathborne+09,ragan+13}. 
These conditions compare well with those present in high-mass star-forming (HMSF) regions
\citep{rathborne+06, simon+06, beuther+s07}.
Follow-up studies confirm IRDCs as birthplaces of high-mass (HM) stars 
and possibly stellar clusters as well \citep{kauffmann+pillai10,henshaw+14,wang+14}. 
Thus, IRDCs offer an opportunity to understand the details of HMSF process, and particularly,
the stellar cluster formation.

Stars form in compact sub-parsec sized cores usually 
embedded at the densest parts of the IRDCs. 
Results of \cite{alves+07} and \cite{andre+10} 
show that the form of the core mass function resembles 
well the Initial Mass Function (IMF) of stars. This
suggests that the process driving the fragmentation
of the progenitor cloud in to cores principally 
establishes the form of the stellar IMF.
Given their compactness, cores can be detected only in
nearby MCs. The densest structures seen in emission
in farther MCs have sizes of $\sim$1~pc and are 
often referred to as clumps. These clumps have mass 
function that resembles the mass function of giant MCs
with an index ranging from 1.8 \cite{kramer+98} to 2.1 \cite{simon+01}.

The similarity in physical properties 
(size, density, mass, etc.) of IRDCs and the clumps \citep{rathborne+06,ragan+13}, 
makes the study the IRDC mass distribution of great importance.  
Studies so far have not found any noticeable differences in the power-law 
($\rm \frac{dN}{dM}\propto M^{\alpha}$) indices for the IRDCs and the clumps
with $\alpha$ clustered around $-2.0$. The results of some of these studies 
are summarized here.
\cite{rathborne+06} reported $\alpha=-2.1$ for clumps embedded in IRDCs, 
whereas \cite{simon+06} using observations in \co\ emission for an IRDCs sample, 
found $\alpha=-1.97$. \cite{ragan+09} reported a broken power law, with $\alpha=-1.76$ 
for masses greater than 40\,\msun.
\cite{peretto+fuller10} (hereafter PF10) reported $\alpha=-1.85$ for masses greater than 
100~\msun\ for IRDCs. 
Recently, \cite{gomez12} using observations at 1.2\,mm, reported $\alpha=-1.6$ 
in the mass range from 15~\msun\ to 3000~\msun, for massive clumps.

Theoretical studies have shown that the HMSF can occur only in regions that have
gas densites above the threshold densities of 
1~g\,cm$^{-3}$ (4790~\msunpc)\citep{krumholz+mckee08}. \cite{rathborne+09}
found that the surface gas densities in IRDCs are often above this threshold density.
\cite{kauffmann+10} defined an empirical mass-size threshold relation to identify 
clouds that may contain HM star formation sites, by studying
Galactic MCs with and without HM star formation. 
\cite{kauffmann+pillai10} found that $\sim$25\% to 50\% of 
the IRDCs in the samples from \cite{rathborne+06} and \cite{peretto+fuller09} 
belong to clouds that are above the mass-size threshold relation.
Star-forming clumps from Bolocam Survey also obey the threshold relation \citep{svoboda+16}.
Thus, there is overwhelming support to the idea that IRDCs are
potential HMSF sites. 

In a recent work, we \citep[hereafter R17]{rr+17}
studied the YSO populations in a sample of 12 Galactic MCs. Each chosen 
MC contains a CS clump associated to young luminous IRAS sources, 
as well as the presence of 
HMSF tracers such
as a methanol, $\rm H_2O$ and OH maser emitting sources and SiO outflows. 
The Line-of-Sight (LoS) to each MC is devoid of
any intervening molecular cloud as inferred from their $^{13}$CO profiles.
We used the Galactic rotation curve model of \cite{clemens85} to infer 
the kinematical distances to all our sample MCs, assigning in all
cases the near distance, when there is distance ambiguity.
The IRDCs catalogued in the present work are embedded in the MCs,
and hence the MC distances also apply to the IRDCs. It is important
to note that this association of our sample IRDCs to MCs eliminates the
usual problem of determining distances to IRDCs. Hence, our distances
to IRDCs only suffer from the error involved in the kinematical distance
method, which is estimated to be 12\% by \cite{faundez+04}.
This distance error is comparable to the 15\% accuracy quoted by
\cite{simon+01} for molecular clouds from the Milky Way Galactic Ring Survey 
using \co~data. 
However, the error in the distance can be much greater 
than the 12\%-15\% mentioned here and vary depending on
longitude and near/far side, as found by \cite{roman-duval09}. 
The introduction of the error on distance is the main source 
of uncertainty on the physical parameters of the IRDCs.

In addition, stellar masses are available for all embedded YSOs. 
All these characteristics of our dataset, offers us an opportunity to study 
quantitatively the high mass star formation, as well as stellar cluster formation 
in IRDCs of wide ranging masses and sizes. The dataset also allows, for the first time, an 
exploration of whether a star formation rate-surface density relation, commonly
referred to as the Kennicutt-Schimdt (K-S) law \citep{kennicutt98},
is applicable within the IRDCs. It may be noted that, K-S law though primarily
applicable on kiloparsec scales in extragalactic objects with a power-law index 
of 1.4, is also applicable within individual Galactic molecular clouds (MCs).
\cite{willis+15} found the index to be varying between 1.77 to 2.86 for HMSF regions, 
whereas for our sample of 12 MCs, index is found to vary between 1.4 and 3.6.

The structure of the paper is the following. In \S2, we describe our sample of IRDCs. Their mass 
and size distributions are given in \S3. The SF properties in our sample of IRDCs are
discussed in \S4. We summarize our conclusions in \S5.

\section{IRDC sample and analysis}

The sample of IRDCs in the present study consists of all IRDCs associated to the 12 
MCs studied by R17, and are drawn from the \cite{peretto+fuller09} IRDC catalog 
(hereafter PF09). All MCs in this sample have at least one HMSF site, as characterized 
by a dense clump detected in CS(J=2--1) line emission and 1.2~mm dust continuum millimeter 
emission \citep{bronfman+96,faundez+04}, associated to maser emitting towards to IRAS 
sources. The MCs were defined using the \co\ (J=1--0) line emission data from the 
Galactic Ring Survey (GRS) \citep{jackson+06}. All the molecular gas that is above a 
column density of gas equivalent to $A_v=1$~mag and is within 5~\kms\ of the 
CS peak velocity is considered to be part of the MC. In Table~1, we give the 
physical properties of these MCs. 
Our MCs are at distances (kinematical) of 1--5.4~kpc, have masses (LTE method) 
in the range 2.3--135$\times10^3$~\msun, and \sigsfr=1--27~\msunmyrpc \citep{rr+17}.

\begin{table*} %[!h!t]
\begin{center}
\centering
\caption{  {Summary of the properties of the MC sample where IRDCs have been searched}}
\begin{tabular}{lccrccccccc}
\hline
\hline
  &    &   &   &      &   &  &  &  &  &  \\
GMC  & IRAS name & Lon & Lat & D     &M$_{\rm LTE}$&\footnotesize N(YSOs) &\footnotesize N(IRDCs) & \siggas\    & \sigsfr\            & SFE   \\
  &      & [deg]     & [deg]   & [kpc]&   [\msun]           &                &                   &\footnotesize [\msunpc] &\footnotesize [\msunmyrpc] &  [\%]   \\
(1) & (2) & (3) & (4)  & (5) & (6)  & (7) & (8) & (9) & (10) & (11) \\
\hline
MC1  & 19230$+$1506  & 50.28 & $-$0.39   & 1.3 (0.2)  & 0.33 (0.17)  &  130  &  6   & 210 (114)  &  3.3 (1.7)  & 2.7 (0.3) \\
MC2  & 19236$+$1456  & 50.22 & $-$0.61   & 3.4 (0.4)  & 3.07 (1.55)  &  556  &  23 & 155 (79)    & 2.1 (1.1)    & 2.2 (0.1) \\
MC9  & 19139$+$1113  & 45.82 & $-$0.28   & 4.8 (0.6)  & 2.56 (1.10)  &  861  &  9   & 646 (332)  & 19.1 (9.6)  & 3.9 (0.2) \\
MC12 & 19132$+$1035  & 45.19 & $-$0.44  & 5.4 (0.6)  & 2.24 (1.12)  & 170  &  3   & 358 (179)   & 1.1 (0.6)  & 0.5 (0.1) \\
MC20 & 19074$+$0814  & 42.43 & $-$0.26  & 4.9 (0.6)  & 2.44 (1.20) &  292  &  5   & 647 (320)   & 6.4 (3.3)  & 1.2 (0.1) \\
MC21 & 19074$+$0752  & 42.11 & $-$0.44  & 3.9 (0.5)  & 3.52 (1.59)  &  672  & 15 & 133 (60)     & 2.3 (1.2)   & 2.6 (0.2) \\
MC23 & 19048$+$0748  & 41.75 &    0.09    & 1.0 (0.1)  & 0.30 (0.22)  &  592  &   1  & 320 (240)  &26.8 (13.5)& 11.9 (0.7)\\
MC75 & 18232$-$1154  & 19.49 &    0.15      & 2.3 (0.3) & 2.32 (1.20)   &  596  &  7  & 613 (320)  & 5.6 (2.8)   & 1.5 (0.1) \\
MC76 & 18236$-$1205  & 19.36 & $-$0.02  & 2.5 (0.3)  & 3.14 (1.74)   &  666  &  25 & 565 (310) & 7.5 (3.7)   & 2.1 (0.1) \\
MC78 & 18223$-$1243  & 18.66 & $-$0.06  & 3.7 (0.4)  & 13.52 (6.46) &  721  &  25 & 763 (365) & 4.6 (2.3)   & 0.8 (0.1) \\
MC80 & 18205$-$1316  & 17.96 & 0.08       & 2.2 (0.3)   & 9.57 (5.03)   &  661  &   7 & 816 (410)  & 5.6 (2.8)   & 1.2 (0.1) \\
MC81 & 18190$-$1414   & 16.94 & $-$0.07 & 2.1 (0.3)  & 0.23 (0.11)   &  14     &  3  & 218 (64)    & 1.0 (0.7)   & 0.8 (0.3) \\

\hline
\end{tabular}

Brief explanation of columns:
(1) Name of the molecular cloud;
(2) Name of the principal IRAS source in the MC;
(3--4) Galactic longitude and latitude in degree;
(5) Kinematical distance (with an error of 12\%) to the dense clump associated to the IRAS source from \citet{faundez+04} ;
(6) Mass of the MC obtained using the LTE method;
(7-8) Number of YSOs and IRDCs spatially associated to the MC;
(9) Surface gas density of the MC;
(10) Surface density of star formation rate of the MC;
(11) Star formation efficiency in \% units in each cloud. 
%}
\end{center}
\end{table*}

\begin{figure}[t]
\centering
\includegraphics[width=1.0\columnwidth]{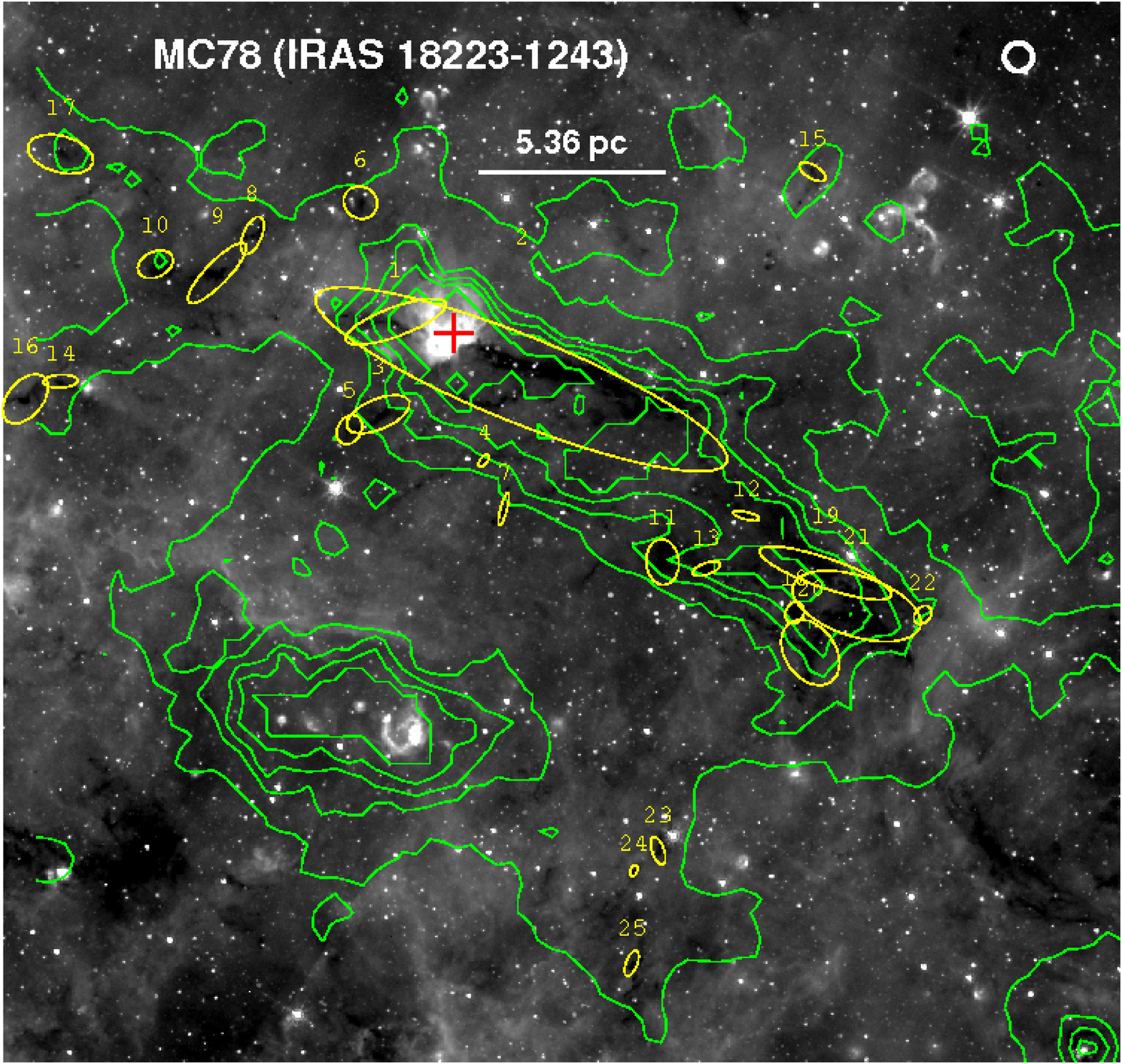}
\caption{\small
Illustration of the locations of our sample of IRDCs (numbered yellow ellipses) on the 
Spitzer 8~\um\ band image (gray-scale) for one of our sample MCs (IRAS 18223-1243; MC78).
The \co\ emission is shown as green contours, starting at
\nhdos$\rm =10^{21}~cm^{-2}$, and increasing in linear steps of $\rm 5.0\times10^{21}~cm^{-2}$.
The MC78 is associated to the protostellar source IRAS 18223-1243, whose location 
is indicated by a red cross. The physical scale of 5.36 pc is equivalent to 5 arcmin angular scale.
{The \co\ beam (white circle) is shown at the top-right corner.}}
\label{fig:fig1}
\end{figure}

\subsection{Selection of IRDCs associated to the molecular clouds}

The PF09 catalog is the most complete source of IRDCs seen on Spitzer 8~\um\ images.
These authors defined an IRDC as a connected structure with a threshold opacity in Spitzer 
8\,\um\ band ($\tau_{8}$) of 0.35 (\nhdos$\rm =2\times10^{21}~cm^{-2}$) and an 
angular size greater than 4\arcsec. 
A visual inspection of our cloud locations on the 8~\um\ Spitzer images shows IRDCs.
We selected all IRDCs from the PF09 catalog that are within the
boundaries of our MCs, defined previously using a columnar gas threshold. 
In the 8 \um\ images of the star-forming (SF) regions of our sample, 
intense emission and absorption zones coexist spatially. This hinders the identification 
of structures such as the IRDCs. Thus, we performed a visual inspection for spatial 
association of the positions of IRDCs and dark patches in the 8~\um\ images. 
The IRDCs without spatial association were rejected from the sample. 
The  column 8 of Table~1 gives the number of IRDCs associated to each MC. 
The complete IRDC sample contains 128 objects; their names, coordinates and 
observed/derived parameters are listed in Table~2. 
The printed version contains 10 sources to guide the reader and the complete table is
given in the electronic version.
The positions of IRDCs most often coincide with the densest \co\ contours of the MCs.
This is illustrated for the IRAS 18223-1243 (MC78) region in Figure~\ref{fig:fig1}.
IRDCs with major axis sizes larger than the \co\ beam size of {46\arcsec\ } 
are clearly associated to the densest contours of the MCs.

\section{Physical properties of the IRDCs associated to the MCs}

In order to obtain the masses of the IRDCs, we obtained the surface gas density 
(\siggas) and physical area of each object. These parameters are obtained using the 
mean opacity ($\tau_8$) in the Spitzer 8~\um\ band and the aparent size of the IRDCs, 
taken from the PF09 catalog. To compute the surface gas density of each IRDC,
we used the expression \citep{butler+tan12},

\begin{equation}
\Big[ \frac{\Sigma_{\rm gas}}{\rm M_{\odot}~pc^{-2}} \Big] =622.7 \times \tau_8,
\end{equation}
{where a gas-to-dust ratio of 156 \citep{ossenkopf+94,butler+tan12}}, 
and an equivalence 
1\,$\rm g\,cm^{-3}$ = 4790 \msunpc\ have been used, assuming a 1 pc depth.

\begin{figure}[!h]
\centering
\includegraphics[width=0.9\columnwidth]{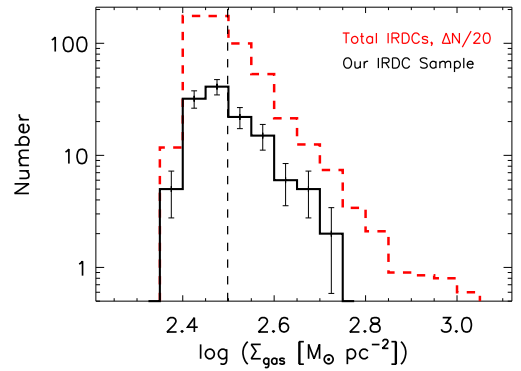}
\caption{\small
Distribution of gas surface density for our sample of IRDCs (black solid histogram) 
compared to that for the full IRDC sample of PF09 (red dashed histogram; after scaling 
down the numbers by a factor of 20). 
The error bars for our sample correspond to the statistical uncertainties in each bin ($\sqrt{N}$). 
The vertical dashed line indicates the {mean value}~(319~\msunpc) for our sample, which
agrees very well with the mean value for the PF09 sample. 
}
\label{fig:fig2}
\end{figure}

The \siggas\ distribution for our sample of IRDCs is shown in Figure~\ref{fig:fig2}, along with that
for the full PF09 sample. The {mean value of 319~\msunpc }
($\sim$0.07~\gcm) for our sample compares very well with the 
mean value for the PF09 sample of \siggas=320~\msunpc. The distributions also have very
similar ranges. This illustrates that our sample of IRDCs is representative of the
full sample of IRDCs in PF09. The properties of our sample %These values 
are also similar to that reported for Galactic dense clumps with or without star formation \citep{schisano+14,ragan+13}.  
Figure~\ref{fig:fig2} shows there is a small range in gas surface densities 
(\siggas$\sim$250 to 500~\msunpc) that likely arises from an observational limitation.
IRDCs are selected in absorption against the galactic background light
at $\sim$8~$\mu$m. This technique is sensitive only for a small range of
gas densities, as is shown in the Fig.~\ref{fig:fig2} for the complete IRDC sample of PF09. 
At low densities, the absorption is not strong enough
to be detected, whereas at high densities, the cloud becomes optically
thick at 8~$\mu$m. The technique underestimates the opacity 
when the region contains an 8~$\mu$m emitting source. When the cloud
starts forming stars, such sources are expected, and hence the true
gas surface densities for star-forming clouds are likely to be higher 
than that infered by $\tau_8$. 
Given that we use the mean value of $\tau_8$ from each IRDC to obtain its 
\siggas~value, an agglomeration of \siggas\ values in a narrow 
dynamic range is expected, as observed in the Figures~\ref{fig:fig2} and \ref{fig:fig9}.

The masses of the IRDCs are obtained using the expression, 
${\rm M}=area\times\Sigma_{\rm gas}$, where the $area$ is the area of the 
ellipse associated to each IRDC (see the definition in PF09) 
and obtained from the major and minor axis angular sizes (see Table~2) and 
distance to the MC
containing each IRDC (Table~1). The effective radius ($\rm R_{eff}$) is 
obtained assuming a circular symmetry and using the expresion, 
$\rm R_{eff} = \sqrt{ area / \pi }$.  
We use this $\rm R_{eff}$ as a size indicator in the analysis of mass-size distribution.
The values of \siggas, $area$, $R_{\rm eff}$, and other 
derived parameters of the clouds are listed in Table~2. 
The uncertainties in the derived parameters are obtained by propagating the distance
errors. 

However, uncertainty in $\tau_8$ can originate from 
several factors (which are considered beyond the scope of this work): 
a) the selection of the MIR opacity model pointed out by \cite{butler+tan12}. 
The variation of opacity per unit gas mass, which is dependent on the model used, 
gives an uncertainty value of up to 30\% in the $\tau_8$, but in IRDCs 
typically is already 20\%. b) The variation on the background intensity, 
which can lead to flux uncertainties of 10\% 
(\siggas $\sim$ 0.01 $\rm g\,cm^{-2}$, for Galactic diffuse MIR emission; 
\cite{butler+tan12}). 
Another source of uncertainty is caused by foreground 
MIR emission along our line of sight to the IRDC. The effect is minimized 
by choosing IRDCs that are relatively nearby, which is the case for the 
IRDCs in the R17 MC sample.

 % ----- muestra de IRDCs -------
 \begin{table*}[t]
 \begin{center}
 \caption{Observed and derivated parameters of individual IRDCs}
 \begin{tabular}{lcrrrrcccrccc}
 \hline
 \hline
 \small ID  & \small Lon   & \small Lat       & a     & b    & \small $\theta$ & $\tau_8$ &\small Area & \small $\rm R_{eff}$ & \small $\rm M$ & $\rm \Sigma$& \sigsfr\ \\

 \small      & \small [deg]                & \small [deg]              &\small ['']&\small [''] & \small [~$^{\circ}$~]  &   & [$\rm pc^2$]  & [pc]              &\small [\msun]          &\tiny [\msunpc]    &\tiny[\msunmyrpc] \\

 \small (1) &\small (2)                       &\small (3)                &\small (4)&\small (5)&\small (6) & \small (7) &    (8)     &         (9)                      &                 (10)       &          (11)        &    (12)\\
 \hline
\small MC78-IRDC1 &  18.6810 &  $-$0.0510 &   87.1  &    23.8  &    20  &  0.60 & 2.07(0.52)  &  0.81(0.06)  &  794(88)         & 383(42)  &  1.69(0.66) \\ 
\small MC78-IRDC2 &  18.6240 &  $-$0.0700 &   361.5&     70.3 & $-$22 &  0.71 & 25.39(6.35)  & 2.84(0.20)   &  11514(1267) & 453(50) & 1.40(0.55)  \\ 
\small MC78-IRDC3 &  18.6890 &  $-$0.0930 &   55.5 &     24.9  &    23  &  0.51 & 1.38(0.35)  &    0.66(0.05) & 450(50)         & 326(36)  & 0.18(0.07)  \\ 
\small MC78-IRDC4 &  18.6410 &  $-$0.1140 &   12.0  &     5.9   &    52  &  0.42 & 0.07(0.02)  &    0.15(0.01) &  19(3)             & 268(30)  & 3.52(1.38)  \\ 
\small MC78-IRDC5 &  18.7020 &  $-$0.1000 &   25.5  &    19.3  &     60 &  0.46 & 0.49(0.12)  &    0.39(0.03) & 144(16)         & 294(32)  & 0.49(0.19)   \\ 
\small MC78-IRDC6 &  18.6970 &  0.0030       &  28.5  &    26.0 & $-$35 &  0.46 & 0.74(0.19)  &    0.48(0.03) & 218(24)         & 294(32)  & 0.32(0.13)  \\ 
\small MC78-IRDC7 &  18.6320 &  $-$0.1360 &   27.7  &     5.3   &   75   & 0.45  & 0.15(0.04)  &    0.21(0.02)  &  42(5)            & 287(32)   & 1.64(0.64)  \\ 
\small MC78-IRDC8 &  18.7460 &  $-$0.0120 &   32.9  &    14.9  &    67  &  0.58 & 0.49(0.12)  &    0.39(0.03) & 181(20)        & 370(41)   & 0.49(0.19)  \\ 
\small MC78-IRDC9 &  18.7620 &  $-$0.0290 &   66.9  &    18.7  &    46  &  0.57 & 1.25(0.31)  &    0.63(0.04)  & 455(51)       & 364(40)   &  0.60(0.23) \\ 
\small MC78-IRDC10 &  18.7900 &  $-$0.0250 &   30.5 &   20.9  &     26 &  0.70 & 0.64(0.16)  &    0.45(0.03)  & 285(32)       & 447(49)  & 5.10(1.98)  \\ 

 \ldots &\ldots &  \ldots &  \ldots &  \ldots & \ldots  &\ldots &  \ldots &  \ldots &  \ldots & \ldots  & \ldots  \\
 
 \hline
 \hline
 \end{tabular}
%\tablecomments{

A brief explanation of the columns: (1) ID of the IRDC; 
(2-3) Galactic coordinates of the IRDC centroid; (4-5) major (a) and minor (b) axes 
of the ellipse that best matches the IRDC; (6) orientation angle of the ellipse; (7) 
(mean) opacity in the Spitzer 8~\um\ band; (8) area of the IRDC; (9) Effective radius 
of the IRDC; 
(10) Mass of the IRDC; (11) Surface gas density, and (12) Surface density of the star formation rate 
of the IRDC.  The data for columns 2--7 are taken of PF09 catalog, and columns 8--12 are derived 
in this work. This table is available in its entirety in the electronic version.
%}
% \end{threeparttable}
 \end{center}
 \end{table*}

\begin{figure}[!h!t]
\centering
\includegraphics[width=1.05\columnwidth]{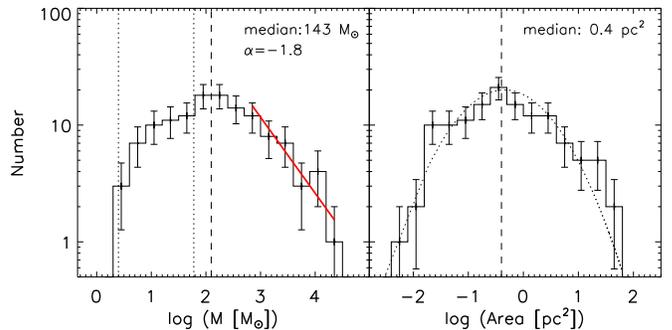}
\caption{\small
Mass (left) and area (right) distribution of the IRDCs. 
The median value of the mass distribution (143~\msun) is shown by the vertical dashed line.
The range of mass for the limit of detection of IRDCs in the PF10 catalog is marked 
by the dotted vertical lines at 2~\msun\ and 50~\msun. The observed distribution for 
higher values of mass than this range is in agreement with a power-law fit (red solid line)
with an index $\sim -1.8$. (Right) The area histogram is well represented
by a log normal function (dashed line). The error bars correspond to the 
statistical uncertainties in each bin ($\sqrt{N}$) in both histograms.}

\label{fig:fig3}
\end{figure}

The distribution of the  mass and area of our IRDCs is shown in Figure~\ref{fig:fig3}.
The vertical dotted lines at 2~\msun\ and 50~\msun\ indicate the limit of
detection in mass, which is obtained using the definition of an IRDC in the 
Spitzer 8~\um\ data (PF10) for a distance range from 1 kpc to 5.4 kpc, 
corresponding to our IRDC sample. For masses above this mass range, 
the distribution follows a power-law functional form given by,

\begin{equation}
\rm \frac{ d~N}{d~M} \propto M^{\alpha},
\end{equation}

\noindent where M is the mass of the IRDC, and N is the 
number of the IRDCs in each mass bin. 
A linear regression (weighted by errors) in this regime 
{gives $\alpha\sim-1.8$}. {This index is consistent with the resulting index of $-$1.85 
of \cite{peretto+fuller10} for IRDCs, and the values $-$2.1, $-$1.97 and $-$1.76 
reported by \cite{rathborne+06,simon+06,ragan+09}, 
respectively, for clump mass distributions in IRDCs.}
It may be noted that a power-law functions with similar index values are
also found for the prestellar and protostellar mass distributions
\citep{konyves+10,andre+10} and for stellar masses in 
young star clusters  (see review of \cite{bastian+10}; \S2.3.2). 

In order to determine the surface density of the star formation rate (\sigsfr) 
in the IRDCs, we obtained the spatial association of the young stellar objects 
(YSOs) within the ellipse defined for each IRDC. The positions and masses 
of the YSOs are taken from R17. 

The method for obtaining the stellar masses and $\Sigma_{\rm SFR}$ 
is fully detailed in R17 and briefly explained below. 
As a first step we determined the bolometric luminosity of each YSO,
using a class-dependent conversion factor from the 24~\um~ luminosity
to bolometric luminosity. We used Spitzer MIR photometry dataset for
determining the SED evolutionary phase (class I, II and III) and the 24~\um~
fluxes. We used the pre-main sequence evolutionary tracks of \cite{tognelli+11} 
to relate bolometric luminosities to stellar masses. We added masses of all YSOs 
within the ellipse defined in columns 4, 5 and 6 of Table 2 to determine the total 
stellar mass associated to each IRDC. We then divided this total mass by the
the area of the ellipse and the
typical timescale that stars spend in each evolutionary class from
\cite{hartmann+kenyon96} and \cite{evans+09} to obtain $\Sigma_{\rm SFR}$. 
For IRDCs without associated YSOs, we assigned a minimum stellar mass 
corresponding to one YSO of mass 0.5 \msun~to obtain a lower
limit value of $\Sigma_{\rm SFR}$. The $\Sigma_{\rm SFR}$ values 
are shown in column 12 of Table 2.

\section{Analysis and discussion}

\subsection{The IRDCs as birth places of high-mass stars}

The high-mass star formation process requires a high mass concentration in a 
relatively small volume. \cite{beuther+07,kauffmann+10,svoboda+16} have 
suggested schemes to study the physical conditions required in the Galactic 
star-forming regions to form HM stars. Particularly, 
\cite{kauffmann+10} (hereafter, K10) found that Galactic low-mass star-forming 
regions satisfy 
the relation $\rm M \leq870~M_{\odot}~\big[\frac{R^{1.33}}{pc}\big]$, where 
$\rm {0.01}<\big[\frac{R}{pc}\big]<{10}$. 
\cite{kauffmann+pillai10} studied the HM star-forming Orion region \citep{hillenbrand97}, 
and established
the presence of massive stars in clouds more massive than that given by the above relation. 
%This criterion has been verified by \cite{kauffmann+pillai10} in the star-forming Orion 
%region, where HMSF activity is in process \citep{hillenbrand97}. For this
%region the K10 relationship is not satisfied. 
For practical purposes, in the present work we redefine the K10 relation as,
\begin{equation}
\rm \Big[~\frac {M}{M_{\odot}} ~\Big]  ~\geq870~\Big[~\frac {R_{eff}}{pc} ~\Big]^{1.33}. 
\end{equation}
In rest of this paper, we will refer this mass-size threshold criterion to form HM stars
as K10 relation, in reference to the \cite{kauffmann+10} paper. These authors 
had also suggested a slightly steeper (index 1.7) relation for a smaller
range of sizes ($\rm 1<\big[\frac{R}{pc}\big]<4$). Considering that our sample covers the full
range of sizes, we use the K10 criterion with index 1.33 in this work.

\begin{figure}[!h]
\centering
\includegraphics[width=1.0\columnwidth]{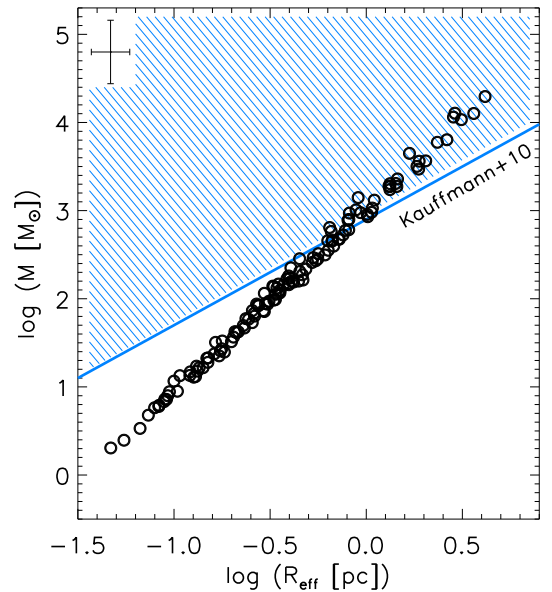}
\caption{\small
Mass-size diagram for the sample IRDCs, using $\rm R_{\rm eff}$ (black circles) 
as proxie for size. The striped area shows the parameter space where IRDCs 
are expected to harbor high-mass star progenitors following the criterion 
of \cite{kauffmann+10} (diagonal solid line).
The mean error of the data is shown in the left-upper region of the diagram.
} 
\label{fig:fig4}
\end{figure}

%A commonly used scheme for selecting potential regions of HM stars is the 
%mass-size diagram \citep{kauffmann+10,kauffmann+pillai10,tan+14}. 
The mass-size diagram for our IRDCs is shown in Figure~\ref{fig:fig4}. In 
this Figure, the K10 criterion is traced with the solid blue line and striped area, while the 
distribution of the masses of the IRDCs as a function of their $\rm R_{\rm eff}$ is plotted 
with the black circles.
The masses and effective radii of the IRDCs were obtained as described in \S3. 
%Rouhgly 37\% (48) of the IRDCs are not associated to YSOs. As mentioned earlier, 
%for these clouds we calculated  lower limit for their star formation
%rate and star formation efficiency corresponding to one YSO of mass of 0.5~\msun. 
A fit was performed on these data, resulting in a power-law relation 
of the form $\rm M = \Sigma_0 \pi R_{\rm eff}^2$. In this relation, the constant $\rm \Sigma_0$ has
a value $\sim$300~\msunpc, consistent with the median surface density value of the IRDCs sample.

In R17, we have tabulated masses of all YSOs embedded in our sample
of IRDCs. In that study, YSOs classified as high
luminosity YSO (HLYSOs; bolometric luminosities $L_{\rm bol}>$10~\lsun) 
are candidates for high mass YSOs. Typically, $L_{\rm bol}$=10~\lsun\ 
corresponds to a star of 1.5\,\msun\ if it is Class I YSO, or
2~\msun\ if it is Class II YSO. Given that YSOs in these early phases are still 
accreting, they are likely to end-up as high mass stars.

\begin{figure}[!h!t]
\centering
\includegraphics[width=1.0\columnwidth]{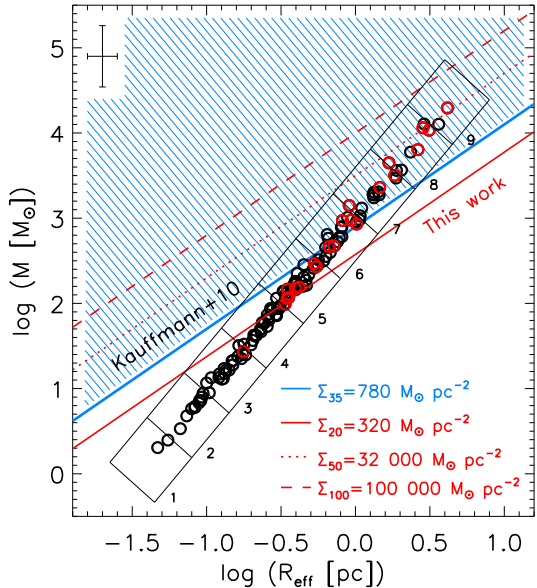}
\caption{\small
Bin distribution of the mass-size IRDC data. Details are the 
same of Figure~\ref{fig:fig4}. The binning is done over the slanted distribution, as is shown with
the numered black boxes. The IRDCs with HLYSOs (defined in the text) are plotted 
with red circles, while the total IRDCs are shown with the black/red circles.
A similar relations to K10 criterion for a threshold HMSF for the IRDCs in this work 
are plotted with the solid, dashed and dotted red lines and are described in the text. 
The mean error of the data is shown in the left-upper region of the diagram.}
\label{fig:fig5}
\end{figure}

\begin{figure} %[!ht]
\centering
\includegraphics[width=0.9\columnwidth]{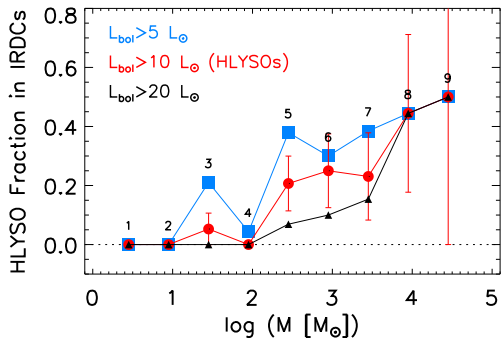}
\caption{\small
HLYSO fraction in IRDCs for each bin defined in the Figure~\ref{fig:fig5} as function of 
masses of the IRDCs, for three ranges of luminosities as is labeled in the 
top-left region. 
}
\label{fig:fig6}
\end{figure}

The availability of masses allows us to analyze the mass-radius relation
for our sample of IRDCs. We started our analysis by identifying
IRDCs that contain at least one HLYSO. These IRDCs are marked in red circles 
in Figure~\ref{fig:fig5}. Clearly there is a tendency for massive-large IRDCs to contain 
a HLYSO, as against lower values of mass-size of IRDCs. However, not all IRDCs satisfying
the K10 criterion harbor a HLYSO. In order to understand the behavior in
more detail, we analyze the fraction of IRDCs containing at least one HLYSO
in small intervals of mass and radius along the mass-radius relation 
followed by our IRDCs. These intervals are shown as boxes in Figure~\ref{fig:fig5}. 
We characterize each box by the mean mass of IRDCs in that box. 
In Figure~\ref{fig:fig6}, we plot the fraction of IRDCs containing at least one HLYSO
as a function of the mean mass for each bin (red line connecting circles). 
In the plot, we also show the fraction of IRDCs that contain at least one
YSO of $L_{\rm bol}>$5~\lsun\ (blue line connecting squares)
and $L_{\rm bol}>$20~\lsun\ (black line connecting triangles).
The fraction of IRDCs containing high luminosity YSOs clearly increases 
with increasing IRDC mass. The frequency
of finding high-mass YSOs in IRDCs of our sample that satisfy the K10 criterion 
is {26\% for $\sim$500~\msun, which increases monotonically to 
$\sim$~50\% for the most massive IRDCs (from 10000 \msun~ to 20000 \msun).}

IRDCs that satisfy the K10 criterion are usually considered as sites of 
massive star formation. 26\% of our IRDCs satisfy the K10 criterion,
but only one third of these contain massive YSOs. 
This suggests that not all IRDCs are currently forming massive stars.
We further find that more massive an IRDC is, it has a higher probability 
of forming a massive star. In our sample, the probability is unity only 
for the IRDCs with mass$>10^4$~\msun. The statistics of the presence or not 
of massive YSOs in our sample of IRDCs allows us to generalize the K10 
mass-size relation, by replacing the ``constant'' in Eq.~3 by $\Sigma_p$:
\begin{equation}
\rm \Big[~\frac {M}{M_{\odot}} ~\Big]  ~\geq\Sigma_p~\Big[~\frac {R_{eff}}{pc} ~\Big]^{1.33}, 
\end{equation}
{where $\Sigma_p$ is the threshold surface gas density of IRDCs that have percentage 
probability $p$ of finding a massive YSO. We find that the K10 criterion 
refers to $p=35$ or $\Sigma_{35}=870$~\msunpc. Our IRDCs have $\Sigma_{20}=320$~\msunpc, 
 $\Sigma_{50}=32~000$~\msunpc.}

\subsection{Star formation efficiency in the IRDCs}

\begin{figure}[!h]
\centering
\includegraphics[width=0.8\columnwidth]{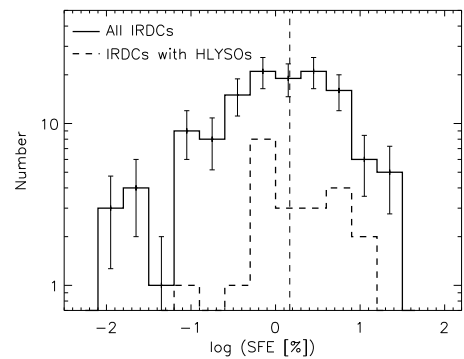}
\caption{\small
SFE Histograms of our IRDC sample. The solid line
shows SFE for our entire sample of 129 objects, whereas the dashed line shows 
the histogram for the IRDCs containing HLYSOs. The vertical line shows the median
value $\sim$1\% for the entire sample.
The error bars correspond to the statistical uncertainties in each bin ($\sqrt{N}$).
}
\label{fig:fig7}
\end{figure}

The star formation activity in the IRDC sample can be measured through the star 
formation efficiency (SFE) and the star formation law. 
For IRDCs, the SFE is obtained  using the simple expression, 
$\rm SFE = M_{\rm star}/(M_{\rm star}+M_{\rm IRDC})$, where $M_{\rm star}$ is the 
``stellar'' mass obtained by adding the masses of the YSOs embedded in each IRDC and  
$M_{\rm IRDC}$ is the mass of the IRDC, obtained in \S3. The masses of the YSOs 
were obtained from the R17 study. 
The SFE distribution (in percentage) for the IRDCs is shown in Figure~\ref{fig:fig7} 
(solid histogram). % , while the dotted histogram show the IRDCs without YSOs. 
%{\color{blue} 

\begin{figure}[!h]
\centering
\includegraphics[width=0.9\columnwidth]{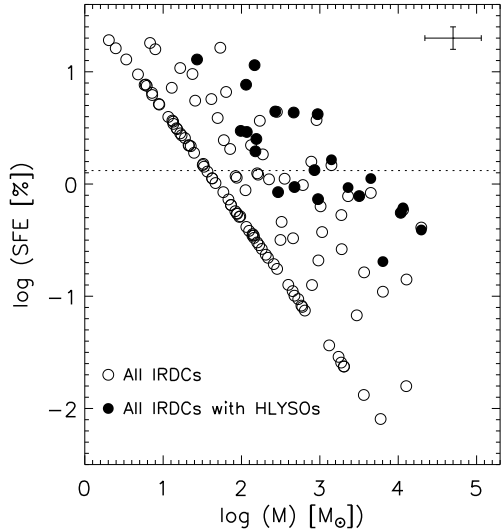}
\caption{\small
SFE$-$Mass distribution of the IRDCs. The open circles show all IRDCs data, while
the filled circles show the position of the IRDCs with HLYSOs associated. The dashed 
line mark the median value of SFE for all IRDCs, $\sim$1\%.  
The mean error of the data is shown in the right-upper region of the diagram.
}
\label{fig:fig8}
\end{figure}

The IRDCs have SFE values between 0.01--30\% with a median value 
of $\sim$1\% (vertical dashed line), whereas the SFE values for IRDCs with HLYSOs 
(plotted with dashed line histogram) range from 0.5--16\%, with similar median value
as for the entire population. Thus, to form an intermediate to high mass star,
the SFE of the IRDC should be higher than 0.5\%. Below this efficiency
IRDCs do not form HLYSOs.

Figure~\ref{fig:fig8} shows the SFE as a function of IRDC mass.
In this figure, the IRDCs harbouring massive stars are distinguished
(solid circles)  from the rest (open circles). 
Above 100\,\msun, IRDCs containing massive stars are the most efficient
in every mass-range. At the same time, there is a systematic decrease in SFE
as the cloud mass increases. Given the mass-radius relation (Figure~\ref{fig:fig4}), 
massive clouds are also larger. Hence, the most efficient HM star forming 
IRDCs are the most compact ones ($R_{\rm eff}<1$~pc), whose mass happens to 
be $<$1000\,\msun\ in IRDC samples.

\subsection{IRDCs and the formation of star clusters}

% {\color{blue}
The definition of star cluster varies widely and depends on our limited 
knoweledge to trace their members and their projected spatial distribution. 
From the literature, some definitions have physical motivation, while others 
are based on their data features. \cite{lada+lada03} define a star cluster 
as groups $\ge$ 35 YSOs with YSO surface densities ($\Sigma_{YSO}$) of 
$\ge$ 3 $\rm YSOs\, pc^{-2}$. In other hand, \cite{carpenter00}, using Ks band 
surface density maps identified clusters as regions with overdensities with 
$\rm \sim 32~YSOs\,pc^{-2}$.  \cite{megeath+12,allen+07} used a similar 
criterion, but with threshold overdensity defined as $\rm \sim 10~YSOs\,pc^{-2}$. 

Our IRDC sample have surface densities ranging from 0.1 to $\sim$
100 $\rm YSOs~pc^{-2}$, with a median value of 6 $\rm YSOs~pc^{-2}$ and 
one outlayer of 160 $\rm YSOs~pc^{-2}$.  Around 40\% of our IRDCs can be 
considered as hosts of
star clusters if we adopt 10 $\rm YSOs~pc^{-2}$ as the threshold
surface density. If the IRDCs with HLYSOs are
only considered, the surface density values vary from
0.1 to 40 $\rm YSOs~pc^{-2}$ , with a median of 2 $\rm YSOs~pc^{-2}$,
which curiously is lower than for our total IRDC sample. 
This trend can be explained as follow, IRDCs with HLYSOs have areas 
greater than $\rm 0.3~pc^{2}$ (median value), as can seen in Figure~\ref{fig:fig3} (right). 
These large clouds generally have longer distances. 
High $\Sigma_{YSO}$ values in these may be due to high numbers of 
HL and LLYSOs detected. Lower $\Sigma_{YSO}$ values may be due 
to few LLYSOs detected, as expected for clouds with far distances. For all 
IRDCs of the sample, the range of $\Sigma_{YSO}$ is greater and 
the median is higher, given the increment of the detection of YSOs 
in nearby clouds, in addtion to the fact that is possible to detect smaller IRDCs, 
achieving a most complete census of their stellar population.  

Additionally, our IRDCs have a \siggas~ ranging from 200 to 400 \msunpc, 
equivalent to 0.04 to 0.08 \gcm~with a mean value of 0.07 \gcm. 
These values are one order of magnitude lower than HM threshold density given 
by \cite{krumholz+mckee08} of 1~\gcm, but similar to values 
founded by \cite{kainulainen+13} in IRDCs and \cite{rathborne+09} for 
Galactic HMSF regions. Thus, our IRDCs could be considered potential sites 
to form star clusters.

\subsection{Star formation law in the IRDCs}

\begin{figure}  %[!ht]
\centering
\includegraphics[width=1.05\columnwidth]{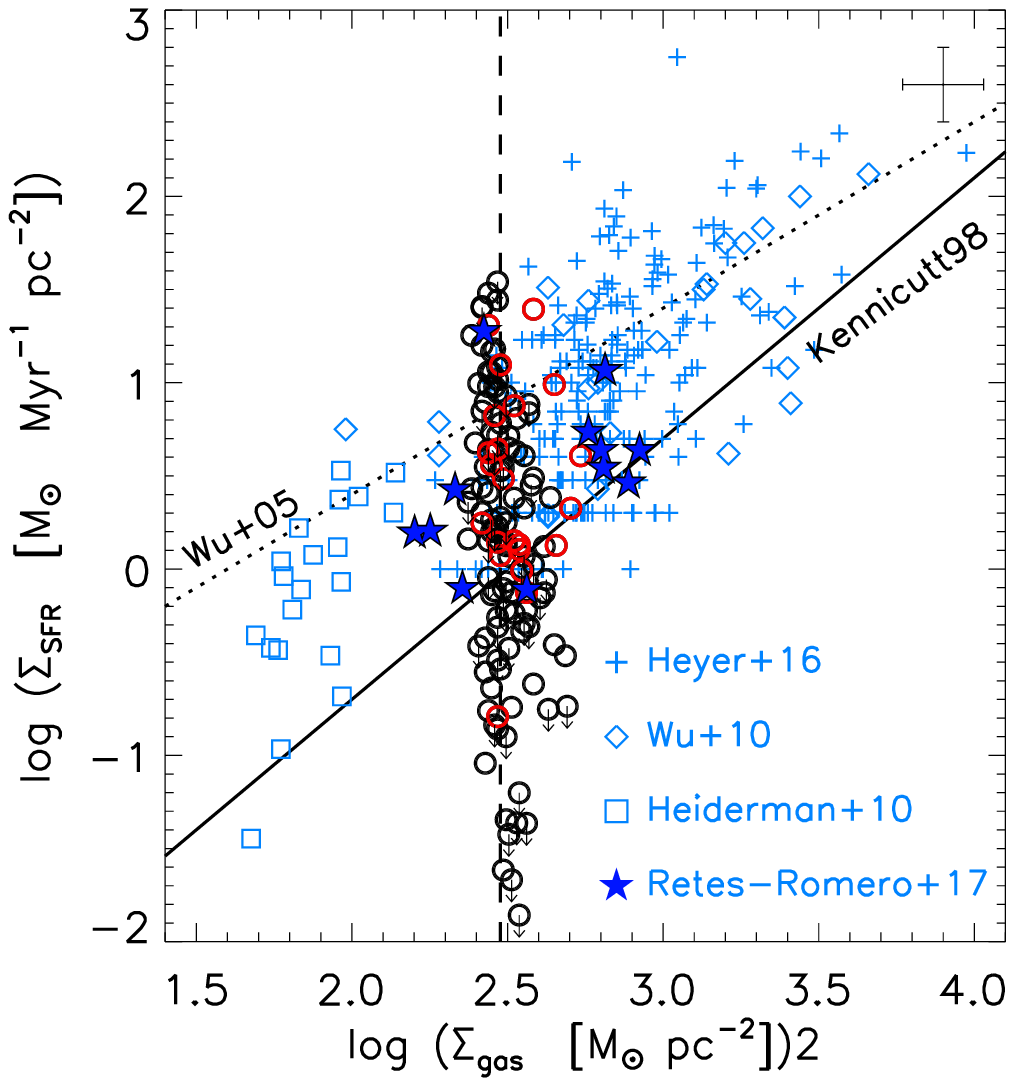}
\caption{\small
\sigsfr\ distribution as function of \siggas\ for the IRDCs. The data for all IRDCs are 
plotted with open black/red circles, whilst the IRDCs with HLYSOs are plotted with open red 
circles, respectively, as in the Figure~\ref{fig:fig5}. The extragalactic relations for this diagram 
are plotted with 
the solid line for the \cite{kennicutt98} and the dotted line for \cite{wu+05} relations. 
In blue diamonds and crosses are plotted the data from Galactic massive clumps, and 
with blue squares are plotted the Galactic nearby LMSF regions (See details in the text). 
In solid blue stars are plotted the Galactic 
HMSF molecular cloud data from R17. The mean error of the IRDC data 
is shown in the upper-right region. 
}
\label{fig:fig9}
\end{figure}

%{\color{black}
In order to study the surface density of the star formation rate (\sigsfr) 
distribution as a function of gas surface density (\siggas) in the IRDCs,
first we obtained the \sigsfr\ using the stellar mass ($\rm M_{star}$) of 
the YSOs embebded in the IRDC. The area used to compute the \sigsfr\ is 
obtained in \S3. The \sigsfr\ and \siggas\ values for the IRDCs 
are shown in the Figure~\ref{fig:fig9}. In this figure, the open circles show 
all IRDCs, with those containing HLYSOs indicated by red circles. 
The IRDCs without YSO detection are plotted with the down-arrow 
symbols in the diagram.
Our sample of IRDCs occupies a very small range of surface gas densities 
around a median \siggas$\sim$319~\msunpc~(vertical dashed line). 
Roughly, 92\% of the IRDCs have \siggas\  values 
between 250~\msunpc\ and 400~\msunpc\ (vertical dotted lines). However, in 
this small range of \siggas, the \sigsfr\ values range over 4 orders of magnitudes 
($\sim10^{-2}~$\msunmyrpc~to~$10^2~$\msunmyrpc). 
%Thus, IRDCs do not showing no relation respect to \siggas\ for the IRDCs.  
The lack of any \siggas-\sigsfr\ relation is in direct contrast with the
well-known relation for star-forming clouds/galaxies established by \cite{wu+05}
(dotted line), and \cite{kennicutt98} (solid line). 

%{\color{blue}
The large spread in \sigsfr\ is likely to be due to the following reasons.
Those IRDCs with \sigsfr\ above the \cite{wu+05} relation are likely to contain
dense gas at small scales that is above the saturation limit at 8\,\um.
Thus, the determined \siggas\ for these IRDCs is highly underestimated.
On the other hand, IRDCs with \sigsfr\ below the Kennicutt (1998) relation  
could be far IRDCs with few locations having the gas surface density 
above the threshold density for star formation, as has been suggested 
in the literature \citep{rr+17,heiderman+10}. 
Additionally, these IRDCs may be in early star formation phase, 
associated to lower \sigsfr values from 0.02 to 5.0 \msunmyrpc, 
or/and their low mass YSO population don't detected due to their large distances
$>$3 kpc (ranging from 3.4 to 5.4 kpc). 

% On the otherhand, IRDCs with \sigsfr\ below the \cite{kennicutt98} relation,
% are large IRDCs with few locations having
% the gas surface density above the threshold density for star formation.
% }

The Figure~\ref{fig:fig9} also shows the global \siggas\ and \sigsfr\ values for the
molecular clouds that habour our sample of IRDCs, taken from R17.
MCs cover a large range in \siggas, but a smaller range of \sigsfr.
In this figure, we also show the \sigsfr-\siggas\ distribution
for Galactic star-forming clumps. Almost all the clumps currently forming 
massive stars from \citet{heyer+16,wu+10} are denser than our IRDCs.
However, our higher \sigsfr\ values are similar to the values for these clumps. 
Compared to nearby SF molecular clouds of \citet{heiderman+10}, our IRDCs 
have higher \siggas\ values by a factor 4, but with similar mean values of \sigsfr. 

One of the characteristics of IRDCs is their small dynamic range in \siggas.
This is likely to arising from the following two effects, the first one 
observational, derived from 
obtaining $\tau_8$ using Spitzer 8\um~data for the IRDCs.
The use of the Galactic MIR background light to obtain $\tau_8$
is sensitive only for a small range of gas densities, then an agglomeration 
of \siggas\ values in a narrow dynamic range is expected, as was discussed earlier 
in section 3.0. The second reason is related to an intrinsic property. 
The fact that \siggas~values to be clustered at
$\sim$300 \msunpc, may be due to the presence of an underlying mass-size relation 
$\rm M\propto L^2$, where L is the maximum linear dimension of the cloud, 
first proposed for molecular clouds by Larson (1981), and more recently 
revised by Lada et al. (2013). 
 Indeed the mass-size relation described in S4.1 give us a 
$\rm M\propto R_{eff}^2$ relation for the IRDCs with a proportionality constant to be 
$\Sigma_0=300$ \msunpc.  Taking into account the above, both can origin the 
narrowing of the \siggas\ range, however the observational bias 
in the selection of the \irdcs~is the most probable cause to produce the clustering 
of \siggas.

\section{Conclusions}

In this work, we studied the properties of IRDCs with and without massive YSOs, and 
the relation between these properties and high-mass star formation.
The IRDCs were searched in molecular clouds harboring embedded 
HM star formation activity. Physical and star formation properties of the IRDCs 
are explored, focusing in the mass-size relation, HMSF threshold and the 
star formation law.  The main conclusions are summarized. In total, 
we have found 835 YSOs in 128 IRDCs. From these, 108 are of intermediate-high mass, 
representing the 13\% of the stellar content embedded in the IRDCs. \\

\begin{itemize}

 \item Our sample of IRDCs have mean surface densities of 319\,\msunpc,
mean mass of 1062~\msun, and a mass function power-law with index $-1.8$, which are 
similar to the corresponding properties for the full sample of IRDCs. The index 
for the mass function is comparable to other studies in IRDCs and Galactic clumps
with high mass star formation.

 \item We find that a 33\% of the IRDC sample contain at least 
one intermediate to massive YSO and satisfy the often-used mass-size criterion to 
forming massive stars. However, not all IRDCs satisfying the mass-size criterion
contain massive stars. whilst that using the presence of HMYSOs in the IRDCs, 
its setting up a  20\% probability for a mass-size relation containing HM young stars.

 \item We find a clear tendency for more massive IRDCs to have a higher probability of
containing a massive YSO. Twenty five (20\%) of our IRDCs are potential sites to harbor 
stellar clusters with masses higher than 100~\msun.

 \item Our sample of IRDCs do not 
show a relation for the star formation law. This is understandable given the 
narrow range of gas surface densities of IRDCs.

\end{itemize}

\acknowledgments
This work was supported by the INAOE research $\rm fellowship$ ``Beca de Colaboraci\'on" granted
to RR. We acknowledge comments and suggestions from an anonymous referee which led to 
refinements in our work and improved the overall presentation of the paper.

\end{document}